\documentclass[graybox, envcountchap]{svmult}

\usepackage{mathptmx}        
\usepackage{amsmath}
\usepackage{amssymb}
\usepackage{mathtools}
\usepackage{color}
\usepackage{helvet}          
\usepackage{courier}         
\usepackage{dirtree}

\usepackage{makeidx}        
\usepackage{graphicx}        
\usepackage{subfig}

\usepackage{multicol}        
\usepackage[bottom]{footmisc}

\usepackage{hyperref}        
\hypersetup{colorlinks=true,urlcolor=blue}

\usepackage[misc]{ifsym}

\newcommand{\LCDM}{$\rm\Lambda CDM$}
\newcommand{\LeDM}{$\rm\Lambda eDM$}

\makeindex             

\begin{document}


\title{Signs of a non-zero equation-of-state\\ for Dark Matter}
\author{Krishna Naidoo}
\institute{Krishna Naidoo (\Letter) \at University College London, Gower Street, London, WC1E 6BT, United Kingdom\\ \email{k.naidoo@ucl.ac.uk}}
%
%
\maketitle

\abstract{We demonstrate how a changing and negative equation-of-state (EoS) for dark matter can alleviate cosmic tensions and explain the integrated Sachs-Wolfe (ISW) void anomaly. We discuss the effect of the model on the cosmic expansion history, growth of structure and the ISW. We show that a negative EoS at late times is able to produce a larger Hubble constant and smaller $\sigma_{8}$, which can explain both cosmological tensions. Furthermore, the model uniquely predicts larger ISW at low redshift, a prediction which is in agreement with observations of larger ISW from voids. The preference for a negative EoS for dark matter at late times is indicative of a unified dark sector and degenerate with models of dark matter and dark energy interaction. Future measurements of the ISW from cosmic voids can provide a unique test for this solution to tensions in cosmology, should they continue to persist.}


\section{Tensions and Anomalies}
\label{sec:1}

The standard model of cosmology \LCDM{} has been extremely successful in describing cosmological observations, including the cosmic microwave background \cite{Planck2020}, galaxy clustering \cite{BOSS2017} and distance ladder \cite{Pantheon2018} measurements. This simple model is comprised of roughly 5\% normal matter, 25\% dark matter and 70\% dark energy. Despite it's many successes the model remains incomplete. Almost 95\% of our current universe is made of stuff we have never directly measured, with few signs that it's properties will be detected in the near future. In \LCDM{} we assume dark matter interacts exclusively through gravity and is cold, meaning it's particles have low velocities that allow them to sink and form large clumps. On the other hand, dark energy is assumed to be a cosmological constant, a scalar field which has a single value across all of space-time. While general relativity allows such a solution, fundamental physics has found explaining its small value extremely difficult. This latter point is often the premise for considering models proposing modifications to gravity.

Over the last decade tensions have emerged between the determination of parameters from the early universe and the late universe. The most egregious of these discrepancies is the Hubble tension, a tension between the expansion rate of the universe today measured from cepheid variables \cite{Riess2022} and those derived from CMB measurements \cite{Planck2020}. Typically measurements from the late universe find a larger Hubble constant than those derived from CMB and early universe physics (including Baryonic Acoustic Oscillations from large scale structure). Combined with the $S_{8}$ tension -- a discrepancy between measurements of the clustering amplitude from the early universe and late-time measurements of weak lensing have placed intense scrutiny on the standard model of cosmology. Are these signs that the simple but powerful assumptions of \LCDM{} are starting to crack, are we seeing symptoms of known limitations or are these just the result of systematics?

A lesser known discrepancy is the Integrated Sachs-Wolfe (ISW) void anomaly. The ISW is a secondary anisotropy which impacts the light from the CMB and is caused by the decay of gravitational potentials, which either draw or impart energy to photons traversing a gravitational potential well (for an overdensity) or trough (for an underdensity). This effect means large scale structure distorts the CMB we observe. There are a number of ways to measure this effect, the simplest approach is to cross-correlate large scale structure observations with the CMB, but this signal is typically quite weak and difficult to measure (see for e.g. \cite{PlanckISW2016}). Another approach is to use stacking, typically carried out with voids (see  \cite{Granett2008,Kovacs2019}). From the latter, cosmologists have measured a surprising anomaly, in that the measurements of void ISW is often significantly larger than what is expected from \LCDM{}. This discrepancy is strongest when measured from photometric voids, which are elongated and preferentially aligned with the line-of-sight. What makes this result particularly puzzling is observations of the lensing signal show no large discrepancies, and therefore no obvious systematics. Instead these measurements are either in agreement with \LCDM{} or slightly lower.

\section{The Dark Matter Solution}
\label{sec:2}

\LCDM{} assumes the presence of cold dark matter, massive uninteractive particles that seed structure formation through their gravitational effects. However, little is known about what dark matter really is; is it a single particle or many, is the microphysics simple or complicated, and how does it fit in with the standard model of particle physics? Its presumed simplicity has proved to be extremely powerful in making predictions but have we reached the limit of these simplistic assumptions, i.e.~has precision cosmology highlighted our ignorance -- could cosmological tensions actually be manifestations of unknown dark matter physics?

Extensions to dark matter often only depart from \LCDM{} on very small scales, when the mass of the dark matter particle starts to become relevant. For example `warm' dark matter models \cite{Colombi1996}, made of moderately light particles, prevent the growth of structures on the very smallest scales due to their larger streaming velocities.
Other extensions such a fuzzy dark matter \cite{Hu2000} similarly depart from \LCDM{} on small scales, but in this  case are caused by extremely light particles with de Broglie wavelengths on kiloparsec scales, causing quantum mechanical interference patterns on the matter distribution field. The tendency for dark matter extensions to be only relevant on small scales has created a perception that dark matter extensions are irrelevant for any solutions to cosmological tensions. However, since we are yet to detect dark matter directly there is plenty of room for unknown physics -- physics that could be relevant on large scales. For example what if dark matter is not a single particle but many, with complicated interactions and decays? What if a subset of the particles are unstable and decay, what if dark matter interacts with other dark sector components -- i.e. dark energy and/or neutrinos? What approach do we take to explore these exotic dark matter properties? One approach is to take each extension and explore them individually, however this is a highly ineffective exploration of the parameter space. An alternative approach is to explore this with phenomenological descriptions. This allows us to explore the parameter space more effectively but comes at the cost of losing specificity, since the dark matter properties will be generally defined. The benefit of such an approach is that we can rely on data to dictate what dark matter models are most viable and relevant for observations. With this in mind we turn to the phenomenological \textit{generalised} dark matter model \cite{Hu1998}. In this model dark matter is modelled as a fluid with free functions for the Equation-of-State (EoS), speed of sound and shear viscosity. Since we are interested in the role this type of model can have on cosmological tensions we will use previous works, in particular \cite{Xu2013,Kopp2018}, to limit the parameters and functional forms considered. In both \cite{Xu2013} and \cite{Kopp2018} a constant EoS for dark matter is constrained extremely tightly to $<10^{-4}$ from CMB experiments. For this reason we will assume at very early times dark matters EoS is effectively zero. While we will assume the speed of sound is adiabatic and shear viscosity is zero.

\subsection{A non-zero Equation-of-State}

With this in mind we will be exploring the consequence of a dark matter model with an EoS that begins at zero at early times and at some scale factor $a_{\rm nz}$ transitions to a monotonic function over cosmic time (specifically, linear as a function of the scale factor $a$);
\begin{equation}
	w_{\rm dm} (a) = \begin{dcases}
		w_{\rm dm,0} \left(\frac{a-a_{\rm nz}}{1-a_{\rm nz}}\right),\quad &\text{for}\, a \ge a_{\rm nz},\\
		0,\quad &\text{otherwise,}\\
	\end{dcases}
\end{equation}
where $w_{\rm dm}(a)$ is the EoS, $w_{\rm dm,0}$ the EoS today, and $a_{\rm nz}$ is the scale factor at which the EoS transitions from zero to non-zero. In \LCDM{} dark matter's EoS is $w_{\rm dm} = 0$. From the continuity equation we can define the time evolution of the density as
\begin{equation}
	\rho_{\rm dm}(a) = \frac{\rho_{\rm dm,0}}{a^{3}}W(a),
\end{equation}
where
\begin{equation}
	W(a) = \exp\left(3\int_{a}^{1}\frac{w_{\rm dm}(a')}{a'}{\rm d}a'\right).
\end{equation}
The total matter distribution is then defined as
\begin{equation}
	\rho_{\rm m}(a) = \frac{\rho_{\rm m,0}}{a^{3}}\Gamma(a),
\end{equation}
\begin{equation}
	\Gamma(a) = f_{\rm b} + f_{\rm \nu} + f_{\rm dm}W(a),
\end{equation}
where $f_{\rm i}$ is the fraction of component $\rm i$ to the matter density today ($\rm b$ for baryons, $\rm \nu$ for non-relativistic neutrinos and $\rm dm$ for dark matter). The addition of the time-varying function $W(a)$ changes the time evolution of gravitational potentials in the Poisson equation, which becomes
\begin{equation}
	\nabla^{2}\Phi(x, a) = \frac{3}{2}H^{2}_{0}a^{2}\Omega_{\rm m}(a)\delta(x, a).
\end{equation}
In the linear regime the time evolution and spatial components of the density contrast can be 
separated to $\delta(x,a) = D(a) \delta(x, 1)$ where $D(a)$ is the linear growth function. This means the time evolution of the gravitational potential is proportional to
\begin{equation}
	\Phi(a) \propto \frac{D(a)}{a}\Gamma(a),
\end{equation}
where as in \LCDM{} $\Gamma(a)=1$. This means the time-derivative of the gravitational potential is given by
\begin{equation}
	\dot{\Phi}(a) = H(a) \left[f(a)-1 + \Upsilon(a)\right] \Phi(a), \label{eq_phidot}
\end{equation}
where $H(a)$ is the Hubble expansion rate, $f(a) = {\rm d} \ln D / {\rm d} \ln a$ the growth rate and 
\begin{equation}
	\Upsilon(a) = \frac{{\rm d}\Gamma(a)}{{\rm d}\ln a} = 3f_{\rm dm}\frac{w_{\rm dm}(a)W(a)}{\Gamma(a)}.
\end{equation}
In \LCDM{}, or any cosmological model where $\Omega_{\rm m}\propto a^{-3}$, this last term ($\Upsilon$) is zero. A non-zero value is a unique signature of the non-zero EoS for dark matter. 

\begin{figure}[h]
	\includegraphics[width=\columnwidth]{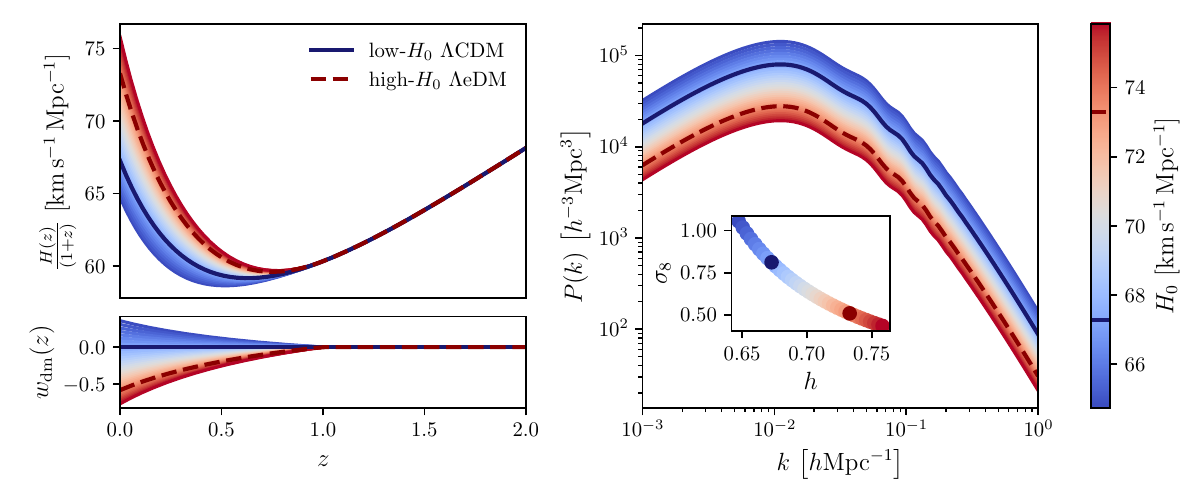}
	\caption{The Hubble function $H(z)$, dark matter EoS $w_{\rm dm}(z)$ and power spectrum $P(k)$ are displayed for models of \LeDM{} as a function of the Hubble constant $H_{0}$. Early universe physics are fixed to CMB constraints. A negative dark mater EoS results in a larger $H_{0}$ and a lower $\sigma_{8}$ (shown in the inset plot on the right). Therefore, \LeDM{} naturally couples the two tensions, providing an explanation for both the $H_{0}$ and $\sigma_{8}$ tensions simultaneously. Profiles for \LCDM{} (assuming early universe CMB constraints) are shown in dark blue (solid lines) and \LeDM{} (assuming late universe constraints on $H_{0}$) are shown in red (dashed lines). The figure was originally presented in \cite{Naidoo2022}.}
	\label{fig:hz_pk}
\end{figure}

\subsection{Impact on the Integrated Sachs-Wolfe and Lensing Signals}

Consider now what effect this will have on the ISW and lensing convergence. The ISW is defined as
\begin{equation}
	\frac{T_{\rm ISW}(\hat{\eta})}{T_{\rm CMB}} = \frac{2}{c^{3}}\int_{0}^{\chi_{\rm LS}} \dot{\Phi}(\chi\hat{\eta},a) a\, {\rm d}\chi,
\end{equation}
and is sensitive to departures from the \LCDM{} prediction that $\Upsilon(a)=0$ (see eq.~\ref{eq_phidot}). While the lensing convergence is defined as
\begin{equation}
	\kappa(\hat{\eta}) = \frac{3H^{2}_{0}}{2c^{2}}\Omega_{\rm m,0}\int_{0}^{\chi_{\rm s}} \frac{\chi}{\chi_{\rm s}}(\chi_{s}-\chi)\frac{D(\chi)}{a(\chi)}\Gamma(\chi)\delta(\chi){\rm d}\chi.
\end{equation}
and is therefore sensitive to departures from the \LCDM{} prediction $\Gamma(a)=1$.

By construction, \LeDM{} is identical to \LCDM{} at early times, since in both cases dark matter EoS is zero, $w_{\rm dm} = 0$. We can use this property to fix the early universe to the conditions of \LCDM{} inferred from Planck \cite{Planck2020}. This allows us to focus on the impact of $w(a > a_{\rm nz}) \neq 0$. In Fig.~\ref{fig:hz_pk} we fix $a_{\rm nz}=0.5$ (equivalent to redshift $z=1$) and vary $w_{\rm dm,0}$ showing a negative EoS for dark matter results in a larger $H_{0}$ and smaller $\sigma_{8}$. The higher $H_{0}$ is the results of a slower decay in the dark matter density, while the lower $\sigma_{8}$ is the result of suppressed structure growth. The coupling of these two effects shows the model can provide an explanation for both cosmological tensions simultaneously.  

\begin{figure}[h]
	\includegraphics[width=\columnwidth]{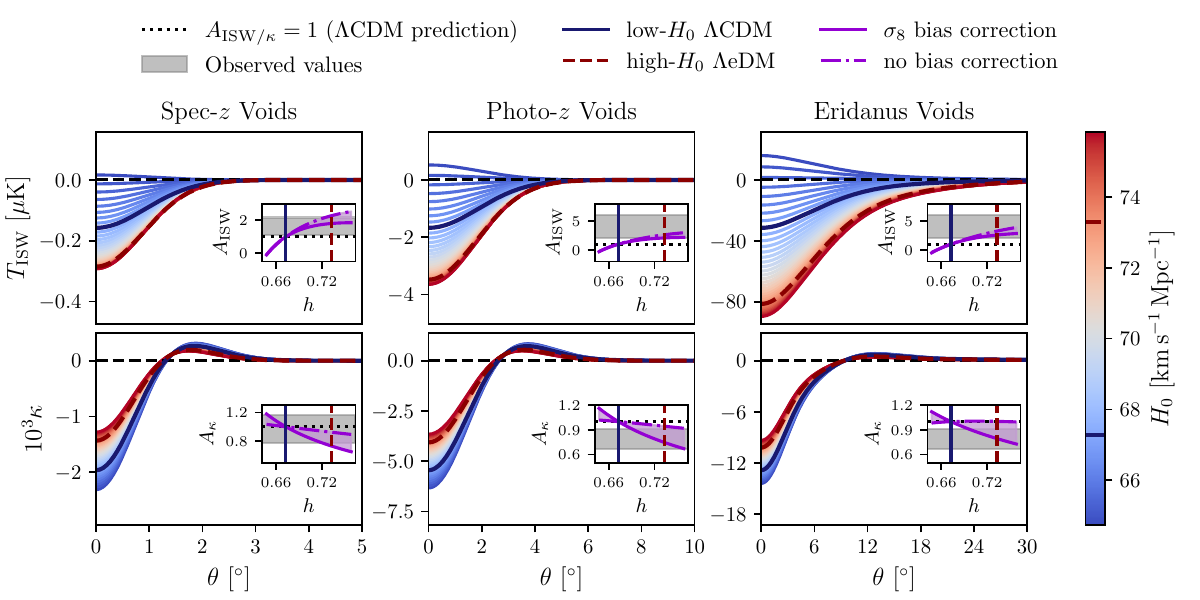}
	\caption{The ISW (top panels) and lensing convergence $\kappa$ (bottom panels) angular profiles for three void types (spec-z on the left, photo-z in the middle and Eridanus voids on the right) are shown as a function of $H_{0}$ in \LeDM{} with all other parameters fixed to early	universe CMB constraints. In the inset plots we show the amplitude of the ISW and $\kappa$ profiles with respect to \LCDM{}, shown with a $\sigma_{8}$ bias correction (solid purple lines) and without a bias correction (dashed purple lines). Measured values are indicated with horizontal grey bands. A higher $H_{0}$ resolves the void-ISW anomaly and recovers lower $\kappa$ amplitudes. For reference the profiles for \LCDM{} assuming early universe CMB constraints are shown with dark blue solid lines and the profiles for \LeDM{} assuming late universe constraints on $H_{0}$ are shown with red dashed lines. The figure was originally presented in \cite{Naidoo2022}.}
	\label{fig:isw}
\end{figure}

In a similar fashion we can explore the effect of \LeDM{} on the ISW and the lensing profiles from voids. We look at three types of voids: spectroscopic voids -- spherical and relatively small \cite{Nadathur2016}, photometric voids -- large and elongated along the line-of-sight \cite{Kovacs2019} and the Eridanus voids -- voids founds along the line-of-sight of the cold spot anomaly \cite{Mackenzie2017} (see \cite{Naidoo2022} for specific void profile properties). In Fig.~\ref{fig:isw} we show the effect of the same models considered in Fig.~\ref{fig:hz_pk}, showing a negative EoS leads to much deeper ISW signals from voids, an effect which is particularly strong for large elongated photometric and Eridanus voids. Rather interestingly this is coupled with equal or slightly lower lensing signals. This is perfectly in keeping with observations, where void ISW signals have been found to be much larger than predicted by \LCDM{} but with lensing signals either lower or in agreement with \LCDM{}. This unique prediction of \LeDM{} is able to explain this long standing anomaly, which makes this model of extreme interest should these anomalies persists in the future.

\subsection{Observational Constraints}

\begin{figure}[h]
	\includegraphics[width=\columnwidth]{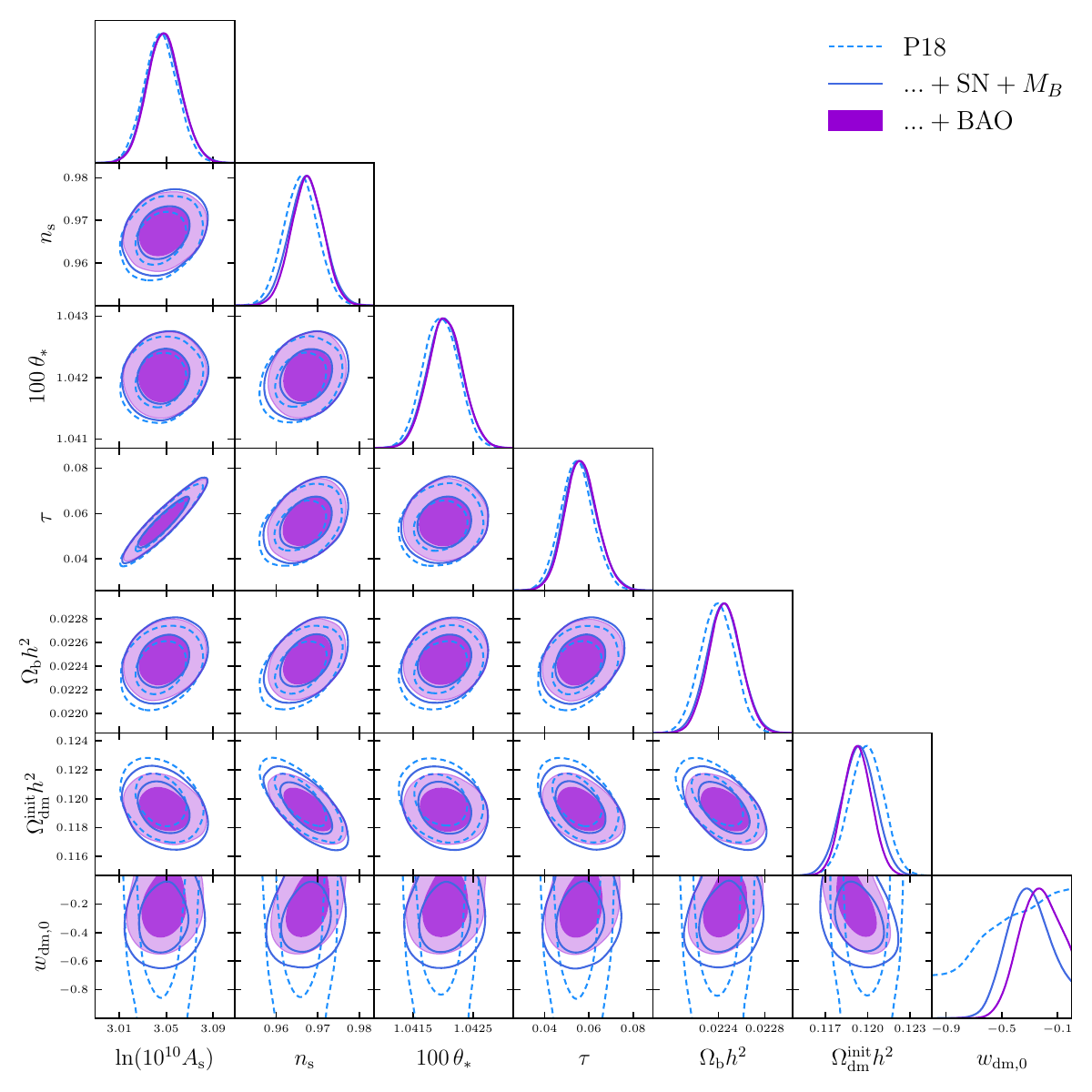}
	\caption{Constraints on the standard 6 parameter cosmological model with the addition of the dark matter EoS today ($w_{\rm dm,0}$). Constraints on the standard 6 parameters are identical to those of \LCDM{}, since by construction \LeDM{} does not alter the early universe. We see that Planck is very unconstraining on the additional $w_{\rm dm,0}$ parameter. The addition of Pantheon supernovae (with SH0ES constraints on the absolute magnitude $M_{B}$) and BAO, significantly constrain $w_{\rm dm,0}$, and while the value is consistent with a zero there is a slight preference for a negative EoS of $w_{\rm dm,0}=-0.2$.}
	\label{fig:params7}
\end{figure}

\begin{figure}[h]
	\sidecaption[t]
	\includegraphics[width=0.6\columnwidth]{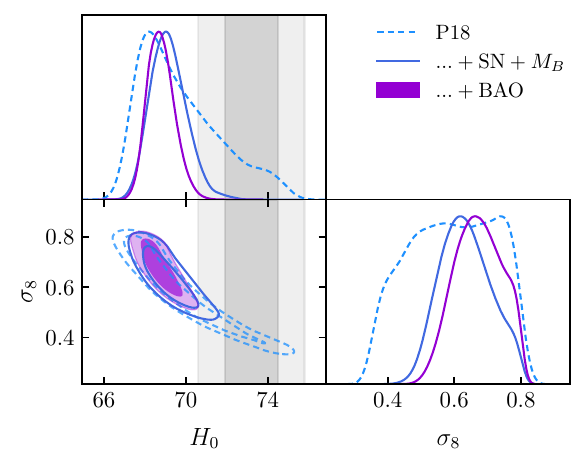}
	\caption{Constraints on \LeDM{} on the Hubble constants $H_{0}$ and $\sigma_{8}$ parameters. This shows a direct anti-correlation between high-$H_{0}$ and low-$\sigma_{8}$.}
	\label{fig:h0_s8}
\end{figure}

In the prior sections we have described how an evolution of the dark matter EoS at late-times can effect cosmological observables, with a keen focus on those relevant to cosmological tensions and the ISW-void anomaly. We can now take this a step further and constrain cosmological parameters with the EoS for dark matter allowed to vary. To avoid exploring largely degenerate parameters we fix $a_{\rm nz}=0.5$ and allow $w_{\rm dm,0}$ to vary between $-1$ and $0$ (the CDM limit). In Fig.~\ref{fig:params7} we show the constraints on the standard 6 parameter model with the addition of $w_{\rm dm,0}$ using observation of the CMB from Planck \cite{Planck2020}, supernova from Pantheon \cite{Pantheon2018} with SH0ES \cite{Riess2022} Cepheid constraints on the absolute magnitude $M_{B}$ and BAO and redshift space distortions from 6dF, SDSS DR7 and BOSS \cite{Beutler2011, Ross2015, BOSS2017}. The constraints on the base \LCDM{} parameters are virtually identical since these are mostly set at early redshift where the models are equivalent. Note, in the case of the dark matter fractional density, we define it according to its asymptotic value at early times, which can be evaluated by calculating
\begin{equation}
	\Omega_{\rm dm}^{\rm init} = \Omega_{\rm dm, 0}\,W(a_{\rm nz}),
\end{equation}
where $\Omega_{\rm dm, 0}$ is the fractional density of dark matter today. This is a measure of what the dark matter fractional density would be today in \LCDM{} and is much easier to sample as this removes a strong degeneracy between $\Omega_{\rm dm, 0}$ and $w_{\rm dm,0}$. With Planck alone $\Omega_{\rm dm, 0}$ is poorly constrained, unsurprisingly since this model considers a late change to \LCDM{}, CMB measurements can only constrain this model through their impact on the sound horizon scale. With the addition of supernovae and BAO the model is better constrained, with SN constraints preferring higher values of $H_{0}$, pushing constraints to higher $w_{\rm dm, 0}$. However, this preference is not definitive and is still consistent with \LCDM{}. In Fig.~\ref{fig:h0_s8} we show the constraints on \LeDM{} for the derived parameters $H_{0}$ and $\sigma_{8}$. This figure shows that there is a strong anti-correlation for high-$H_{0}$ and low-$\sigma_{8}$. In other words a small increase in $H_{0}$ is balanced with a significantly larger shift in $\sigma_{8}$. So while this model does correlate both tensions, raising $H_{0}>70\,{\rm Km\,s^{-1}\,Mpc^{-1}}$, as measured by \cite{Riess2022}, are strongly disfavoured due to the resulting low $\sigma_{8}$.

A negative EoS for dark matter at late-time is a promising solution to tensions and the ISW-void anomaly. Currently, cosmological data shows a marginal preference for \LCDM{} \cite{Naidoo2022}, with more precise measurements from void ISW and lensing (or more typical galaxy and weak lensing auto and cross-correlations) required to make a more definitive statement. Current best-fit constraints on \LeDM{}, with a negative EoS for dark matter at late-times, predict a dampening of the void lensing signal by a factor of $\sim 0.9$  with respect to \LCDM{} and an increase in the void ISW signal by a factor of $\sim 2$ with respect to \LCDM{}. These predictions fit perfectly with observational measurements from voids. To our knowledge, this is the only model that can simultaneously provide a solution to the Hubble and $S_{8}$ tension, and the void-ISW anomaly.

\subsection{Implications: an interacting Dark Sector}

A negative EoS for dark matter is a significant departure from the cold dark matter paradigm. So what does a negative EoS for dark matter actually mean? Generally, the EoS for the different constituents in the universe are as follows, radiation has an EoS = 1/3, cold dark matter and baryons have an EoS = 0 and a cosmological constant has an EoS = -1. Imagine now that dark matter was unstable and decayed into radiation, over time its \emph{effective} EoS would be slightly positive, since it would describe a combination of dark matter and radiation decay products. Similarly, a slightly negative EoS can be thought of as being an effective description of dark matter that decays or interacts with dark energy. For this reason the model is degenerate with models of interacting dark matter and dark energy \cite{Farrar2004}. To follow the behaviour of \LeDM{}, the interactions would need to be relevant only at very late-times. This could be achieved if this interaction/decay only occurred at low energies or densities, or if the dark matter particle was unstable and decayed with a very long half-life. This behaviour is similar to the properties of the evanescent matter model of  \cite{Peebles2012}. Should this turn out to be true, this would mark a significant departure from \LCDM{} and imply that dark energy is not a cosmological constant but rather something else. Furthermore it would mean that cosmology could provide a window into new physics in the dark sector, physics which has remained elusive to direct detection from particle physics experiments.

\section{Future work and predictions for experiment}
\label{sec:4}

A non-zero EoS for dark matter can alleviate tensions in cosmology and provide an explanation for the ISW-void anomaly. Whether this models holds up against the test of time depends on a number of factors: (1) that tensions and the ISW-void anomaly remain, without clear systematics in observations and/or methodology being discovered and (2) that limitations to the model are resolved.

Generalised dark matter is a good way to explore exotic dark matter models, however the model is limited to linear perturbation theory and has not been explored in great detail in high-resolution simulations and on small scales. Exploring these model in simulations will allow us to test whether such a model is compatible with observations and demonstrate how these models behave on small scales.

A unique consequence of this model is its prediction of larger ISW at low redshift. Improved ISW measurements from voids, including better understanding of void galaxy bias, will help to clearly establish the strength of the void-ISW anomaly -- it's persistent will be the clearest sign of a negative EoS for dark matter at late times.

\end{document}